\documentclass[10pt, conference]{IEEEtran}

\DeclareMathAlphabet{\mathpzc}{OT1}{pzc}{m}{it}

\newtheorem{theorem}{Theorem}

\newcommand{\qed}{\nobreak \ifvmode \relax \else
      \ifdim\lastskip<1.5em \hskip-\lastskip
      \hskip1.5em plus0em minus0.5em \fi \nobreak
      \vrule height0.75em width0.5em depth0.25em\fi}

\usepackage{psfrag, epsfig, amsfonts, algorithm, algorithmic}
\usepackage{cite, graphicx, psfrag, subfigure, url, amsmath, amssymb}

\usepackage{graphicx}
\usepackage{epstopdf}
\begin{document}
%
\title{Spectrum Sensing by Cognitive Radios at Very Low SNR}
%


\author{\authorblockN{Zhi Quan$^{1}$, Stephen J. Shellhammer$^1$, Wenyi Zhang$^1$, and Ali H. Sayed$^2$}
\authorblockA{
$^1$Qualcomm Incorporated, 5665 Morehouse Drive, San Diego, CA
92121 \\
E-mails: \{zquan, sshellha, wenyiz\}@qualcomm.com \\
$^2$Electrical
Engineering Department, University of
California, Los Angeles, CA 90095\\
E-mail: sayed@ee.ucla.edu }\vspace{-10pt} }

\markboth{}{}

\maketitle

\begin{abstract}

Spectrum sensing is one of the enabling functionalities for
cognitive radio (CR) systems to operate in the spectrum white
space. To protect the primary incumbent users from interference,
the CR is required to detect incumbent signals at very low
signal-to-noise ratio (SNR). In this paper, we present a spectrum
sensing technique based on correlating spectra for detection of
television (TV) broadcasting signals. The basic strategy is to
correlate the periodogram of the received signal with the \emph{a
priori} known spectral features of the primary signal. We show
that according to the Neyman-Pearson criterion, this spectra
correlation-based sensing technique is asymptotically optimal at
very low SNR and with a large sensing time. From the system design
perspective, we analyze the effect of the spectral features on the
spectrum sensing performance. Through the optimization analysis,
we obtain useful insights on how to choose effective spectral
features to achieve reliable sensing. Simulation results
show that the proposed sensing technique can reliably detect
analog and digital TV signals at SNR as low as $-20$ dB.
\end{abstract}


\IEEEpeerreviewmaketitle

\section{Introduction}

Due to the increasing proliferation of wireless devices and
services, the traditional static spectrum allocation policy
becomes inefficient. The Federal Communications Commission (FCC)
has recently opened the TV bands for cognitive radio devices,
which can continuously sense the spectral environment, dynamically
identify unused spectral segments, and then operate in these white
spaces without causing harmful interference to the incumbent
communication services \cite{FCC2008}. The IEEE 802.22 Wireless
Regional Area Network (WRAN) working group is developing a
CR-based air interface standard for unlicensed operation in the
unused TV bands \cite{Shellhammer2006}.

Spectrum sensing to detect the presence of primary signals is one
of the most important functionalities of CRs. To avoid causing
harmful interference to the incumbent users, FCC requires that
unlicensed CR devices operating in the unused TV bands detect TV
and wireless microphone signals at a power level of $-114$ dBm
\cite{FCC2008}. For a noise floor around $-96$ dBm in the receiver
circuitry (with respect to 6 MHz bandwidth and a 10 dB noise
figure), spectrum sensing algorithms need to reliably detect
incumbent TV signals at a very low SNR of at least $-18$ dB. This
requirement poses new challenges to the design of CR systems since
traditional detection techniques such as energy detection and
matched filtering are no longer applicable in the very low SNR
region \cite{Shellhammer2006}.

In general, there are three signal detection approaches for
spectrum sensing: energy detection, matched filtering (coherent
detection), and feature detection. If only the local noise power
is known, the energy detector is optimal \cite{Kay98II}. If a
deterministic pattern (e.g., pilot, preamble, or training
sequence) of primary signals is known, then the optimal detector
usually applies a matched filtering structure to maximize the
probability of detection. Depending on the available \emph{a
priori} information about the primary signal, one may choose one
of the above approaches for spectrum sensing in CR networks.
However, energy detection and matched filtering approaches are not
applicable to detecting weak signals at very low SNR. At very low
SNR, the energy detector suffers from noise uncertainty and the
matched filter experiences the problem of lost synchronization. To
improve sensing reliability, most previous studies have focused on
the development of cooperative sensing schemes using multiple CRs
\cite{Quan08JSTSP}\cite{Quan08SPM}\cite{Quan2009wb}. An
alternative approach is to use feature detection provided that
some information is \emph{a priori} known. Cyclostationary
detection exploiting the periodicity in the modulated schemes
\cite{Gardner91} is such an example but requires high
computational complexity. Recently, Zeng and Liang developed an
eigenvalue based algorithm using the ratio of the maximum and
minimum eigenvalues of the sample covariance matrix
\cite{Zeng2008}.

In this paper, we develop a feature detection-based spectrum
sensing technique for a single CR to meet the FCC sensing
requirement. The basic strategy is to correlate the periodogram of
the received signal with the selected spectral features of a
particular TV transmission scheme, either the national television
system committee (NTSC) scheme or the advanced television standard
committee (ATSC) scheme, and then to examine the correlation for
decision making. By utilizing the asymptotic properties of
Toeplitz matrices \cite{Gray2006}, we show that for certain signal models the spectra
correlation-based detector is asymptotically equivalent to the
likelihood ratio test (LRT) at very low SNR. In addition, we
analyze how the spectral features can affect the sensing
performance. Specifically, we formulate the sensing problem into an optimization problem. By solving this problem, we obtain
useful insights on how to select or design effective spectral
features to achieve reliable sensing. Extensive simulation results
show that the proposed sensing technique can reliably detect TV
signals from additive white Gaussian noise (AWGN) at SNR as low
as $-20$ dB.


\section{Spectra Correlation Based Spectrum Sensing}\label{sec:sensing}

Before presenting the spectrum sensing technique, we first briefly
review the TV transmission schemes.

\subsection{TV Signal Characteristics}

\begin{figure}[t]
\centering \subfigure[The measured NTSC channel spectrum in
UHF Channel 51 (San Diego, CA, USA).]{\epsfig{file=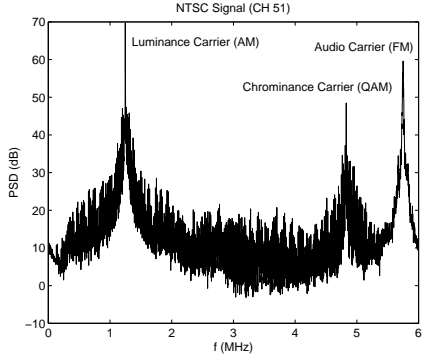,width=2.5in}} \subfigure[The
measured ATSC channel spectrum in UHF Channel
19 (San Diego, CA, USA).]{\epsfig{file=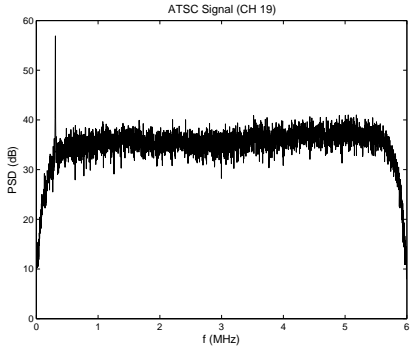,width=2.5in}} \caption{The estimated
power spectra in NTSC and ATSC channels.}\label{fig:PSD}
\end{figure}

A typical TV channel occupies a total bandwidth of $6$ MHz and its
power spectrum density (PSD) describes how the signal power is
distributed in the frequency domain. Fig. \ref{fig:PSD} (a) and
(b) illustrate the PSD functions of both NTSC and ATSC signals.
NTSC is the standardized analog video system used in North America
and most of South America. The power spectrum of an NTSC signal
consists of three peaks across the $6$ MHz channel, which
correspond to the video, color, and audio carriers, respectively.
On the other hand, ATSC is designed for the digital television
(DTV) transmission, and it delivers a Moving Picture Experts Group
(MPEG)-2 video stream of up to $19.39$ Mbps. The ATSC spectrum is
relatively flat but has a pilot located in $310$ kHz above the
lower edge of the channel.

We find that both NTSC and ATSC signals have distinct spectral
features, which are constant during the transmissions. This
observation motivates us to design a spectrum sensing technique
for TV signals by exploiting these \emph{a priori} known spectral
features.

\subsection{Sensing Strategy}


The spectrum sensing problem can be modeled into a binary
hypothesis test at the $l$-th time instant as follows:
\begin{equation}\label{eqn:binary-hypotheses}
\begin{array}{ll}
\mathcal{H}_0:~ y(l)= v(l), \ \ l=0, 1, 2, \ldots;\\
\mathcal{H}_1:~ y(l)= x(l) + v(l), \ \ l=0, 1, 2, \ldots,
\end{array}%
\end{equation}
where $y(l)$ is the received signal by a secondary user, $x(l)$
denotes the transmitted incumbent signal, and $v(l)$ is assumed to
be complex zero-mean additive white Gaussian noise (AWGN), i.e.,
$v(l) \sim \mathcal{CN} (0, \sigma_v^2)$. We assume that the
signal and noise are independent. Accordingly, the PSD of
the received signal $S_{Y}(\omega)$ for different hypotheses can
be written as
\begin{equation}
\begin{array}{ll}
     \mathcal{H}_0: &  S_{Y}(\omega) =   \sigma_v^2    \\
      \mathcal{H}_1: & S_{Y}(\omega) = S_{X}(\omega) + \sigma_v^2, \ \ 0\leq \omega < 2 \pi,  \\
\end{array}%
\end{equation}
where $S_{X}(\omega)$ is the PSD function of the transmitted
primary signal. Our objective is to distinguish between
$\mathcal{H}_0$ and $\mathcal{H}_1$ by exploiting the unique
spectral signature exhibited in $S_{X}(\omega)$.

Generally, we can obtain an estimate of the PSD of the observations through various spectral estimation algorithms, and here we focus on the periodogram, {\it i.e.}, the squared magnitudes of the $n$-point discrete-time Fourier
transform (DFT) of the $n$-point received
signal, denoted
\begin{equation}
S_{Y}^{(n)}(k), \ \ k=0, 1, \ldots, n-1.
\end{equation}
On the other hand, we suppose that the $n$-point sampled PSD of the signal under detection, $S_X^{(n)}(k) = S_X(2\pi k/n)$,
is known \emph{a priori} at the receiver. To detect the presence
of a TV (NTSC or ATSC) signal, we perform the following test:
\begin{equation}\label{eqn:detector}
T_n = \frac{1}{n} \sum_{k=0}^{n-1} S_Y^{(n)}(k) S_X^{(n)}(k) \begin{array}{c} \mathcal{H}_{1} \\ \gtreqless \\
\mathcal{H}_{0} \end{array}  \gamma
\end{equation}
where $\gamma$ is the decision threshold. Namely, if the spectra
correlation between $S_X^{(n)}(k)$ and $S_Y^{(n)}(k)$ is greater
than the threshold then we would decide $\mathcal{H}_1$, i.e.,
presence of the signal of interest; otherwise, we would decide
$\mathcal{H}_0$, i.e., absence of the primary signal.

\section{Asymptotic Optimality}\label{sec:LRT}

In this section, we show that the proposed spectrum sensing
technique (\ref{eqn:detector}) is asymptotically optimal at very
low SNR in the Neyman-Pearson sense. The asymptotic optimality is in the sense that, as shown in Theorem 1 below, the decision statistic $T_n$ asymptotically approaches the likelihood ratio decision statistic for low SNR and large observation length.

\subsection{LRT at Very Low SNR}

Considering a sensing interval of $n$ samples, we can represent
the received signal and the primary transmitted signal in vector
form as $\mathbf{y} = \left[y(0), y(1), \ldots, y(n-1) \right]^T$
and $\mathbf{x} = \left[x(0), x(1), \ldots, x(n-1) \right]^T$.
Since TV signals are perturbed by propagation along multiple paths, it may be reasonable to
approximately model them as being a second-order stationary
zero-mean Gaussian stochastic process, i.e,
\begin{equation}
\mathbf{x} \sim \mathcal{CN}\left(\mathbf{0},
\boldsymbol{\Sigma}_n \right)
\end{equation}
where
\begin{equation}
\boldsymbol{\Sigma}_n  = \mathbb{E}\left(\mathbf{x}\mathbf{x}^T
\right)
\end{equation}
is the covariance matrix. Consequently,
(\ref{eqn:binary-hypotheses}) is equivalent to the following
hypothesis testing problem in the $n$-dimensional complex space
$\mathcal{C}^{n}$:
\begin{equation}
\begin{array}{ll}
     \mathcal{H}_0: &  \mathbf{y} \sim \mathcal{CN}\left(\mathbf{0}, \sigma_v^2 \mathbf{I}\right)     \\
      \mathcal{H}_1: & \mathbf{y} \sim \mathcal{CN}\left(\mathbf{0}, \boldsymbol{\Sigma}_n+\sigma_v^2 \mathbf{I}\right)
\end{array}%
\end{equation}
where $\mathbf{I}$ is the identity matrix. The logarithm of the likelihood
ratio is given by \cite{Poor1994}:
\begin{equation}
\begin{split}
\log L(\mathbf{y}) = &2n \log \sigma_v - \log
 \mathrm{det}\left( \boldsymbol{\Sigma}_n + \sigma_v^2 \mathbf{I}\right)  \\
 &- \mathbf{y}^T \left[ \left(\boldsymbol{\Sigma}_n+ \sigma_v^2 \mathbf{I} \right)^{-1} - \sigma_v^{-2} \mathbf{I} \right] \mathbf{y}
\end{split}
\end{equation}
Incorporating the constant terms into the threshold, we obtain the
logarithmic LRT detector in the quadratic form as
\begin{equation}\label{eqn:LRT}
T_{\mathrm{LRT}} =  \mathbf{y}^T \left[\sigma_v^{-2} \mathbf{I}
-\left(\sigma_v^2 \mathbf{I} + \boldsymbol{\Sigma}_n \right)^{-1}
\right] \mathbf{y} \begin{array}{c} \mathcal{H}_{1} \\ \gtreqless \\
\mathcal{H}_{0} \end{array}  \gamma'
\end{equation}
which is the optimal detection scheme according to the
Neyman-Pearson criterion. This detector is also known as a
\emph{quadratic detector}.

From the Taylor series expansion, we have
\begin{align}\label{eqn:expansion}
\left(\sigma_v^2 \mathbf{I} + \boldsymbol{\Sigma}_n \right)^{-1}
&=\left(\mathbf{I} + \sigma_v^{-2} \boldsymbol{\Sigma}_n \right)^{-1} \sigma_v^{-2}  \nonumber \\
&=\left( \mathbf{I} -\sigma_v^{-2} \boldsymbol{\Sigma}_n +
\sigma_v^{-4} \boldsymbol{\Sigma}_n^{2} -\cdots  \right) \sigma_v^{-2} 
\end{align}
where the convergence of the series is obtained if the eigenvalues
of $\sigma_v^{-2} \boldsymbol{\Sigma}_n$ are less than unity. This condition always holds in the low SNR regime where $\sigma_v^2$ grows sufficiently large. For weak signal detection in the very low SNR
region, i.e.,
$\mathrm{det}^{1/n}\left(\boldsymbol{\Sigma}_n\right) \ll
\sigma_v^{2}$, (\ref{eqn:expansion}) can be approximated as
\begin{equation}\label{eqn:appr}
\left(\sigma_v^2 \mathbf{I} + \boldsymbol{\Sigma}_n \right)^{-1}
\simeq \sigma_v^{-2} \mathbf{I}-\sigma_v^{-4}
\boldsymbol{\Sigma}_n
\end{equation}
Plugging (\ref{eqn:appr}) into (\ref{eqn:LRT}), we obtain
\begin{equation}\label{eqn:LRTlowSNR}
T_{\mathrm{LRT}} =  \mathbf{y}^T \left[\sigma_v^{-2} \mathbf{I}
-\left(\sigma_v^2 \mathbf{I} + \boldsymbol{\Sigma}_n \right)^{-1}
\right] \mathbf{y} \nonumber \\
\simeq \sigma_v^{-4} \mathbf{y}^T \boldsymbol{\Sigma}_n \mathbf{y}
\end{equation}
Hence, the optimal LRT detector at very low SNR is given by
\begin{equation}\label{eqn:LRT_lowSNR}
T_{\mathrm{LRT},n} \simeq \frac{1}{n} \mathbf{y}^T \boldsymbol{\Sigma}_n \mathbf{y}  \begin{array}{c} \mathcal{H}_{1} \\ \gtreqless \\
\mathcal{H}_{0} \end{array}  \gamma_{\mathrm{LRT}}
\end{equation}
where $\gamma_{\mathrm{LRT}}= \sigma_v^4 \gamma'/n$.


\subsection{Asymptotic Equivalence}

Now we show that our proposed spectra correlation-based detector
(\ref{eqn:detector}) is asymptotically equivalent to the LRT
detector at very low SNR (\ref{eqn:LRT_lowSNR}). Consider a
sequence of optimal LRT detectors as defined in
(\ref{eqn:LRT_lowSNR})
\begin{equation}\label{eqn:LRT_detector_seq}
T_{\mathrm{LRT}, n} = \frac{1}{n} \mathbf{y}^T \boldsymbol{\Sigma}_n \mathbf{y}  \begin{array}{c} \mathcal{H}_{1} \\ \gtreqless \\
\mathcal{H}_{0} \end{array}  \gamma_{\mathrm{LRT}},\ \ n=1, 2,
\ldots.
\end{equation}
Likewise, we define a sequence of spectra correlation detectors
as
\begin{equation}\label{eqn:SC_detector_seq}
T_n =\frac{1}{n} \sum_{k=0}^{n-1} S_X^{(n)}(k) S_Y^{(n)}(k), \ \
n=1, 2, \ldots.
\end{equation}
Note that the LRT detectors are performed in time domain while the
spectra correlation detectors are in frequency domain. The
asymptotic equivalence of these two sequences of detectors is
established in the following theorem.
\begin{theorem}\label{thm1}
The sequence of spectra correlation detectors $\{T_n\}$ defined
in (\ref{eqn:SC_detector_seq}) are asymptotically equivalent to
the sequence of optimal LRT detectors $\{T_{\mathrm{LRT}, n}\}$ at
very low SNR defined in (\ref{eqn:LRT_detector_seq}), i.e.,
\begin{equation}
\label{eqn:theorem1}
\lim_{n\rightarrow \infty} \left|T_{\mathrm{LRT}, n}-T_n
\right|=0.
\end{equation}
\end{theorem}
\vspace{6pt}
\begin{proof}
The proof is sketched in Appendix \ref{appendix:a}.
\end{proof}

\section{Spectral Feature Selection} \label{sec:opt}

In this section, we study the effect of spectral features on the
detection performance. Although the sensing algorithm cannot
control or change the spectral features of transmitted signals
since these features are completely determined by the incumbent
transmitter, we can obtain through analysis important insights to
identify the best features for the signal detection. These
insights are also useful for system engineers to design robust
signals that can be reliably detected at very low SNR.



We first consider the case where there is no primary signal in the
band of interest. Under hypothesis $\mathcal{H}_0$, we have
\begin{align}
\mathbb{E}[T_{n,0}] =
 \frac{1}{n}  \sigma_v^2 \sum_{k=0}^{n-1} S_X^{(n)}(k)   = \sigma_v^2 P_x
\end{align}
where
\begin{equation}
P_x = \frac{1}{n} \sum_{k=0}^{n-1} S_X^{(n)}(k)
\end{equation}
is the average power transmitted across the whole bandwidth. On
the other hand, by exploiting the fact that the periodogram is an asymptotically unbiased estimate of the PSD \cite{stoica97:book}, we have for sufficiently large $n$,
\begin{align}
 \mathbb{E}[T_{n,1}] &=
\frac{1}{n} \sum_{k=0}^{n-1}
\mathbb{E}[S_Y^{(n)}(k)] S_X^{(n)}(k) \nonumber \\
&\approx \sigma_v^2 P_x +  \frac{1}{n} \sum_{k=0}^{n-1} \left[S_X^{(n)}(k)\right]^2
\end{align}
under hypothesis $\mathcal{H}_1$.

\begin{figure}[t]
\centering{\epsfig{figure=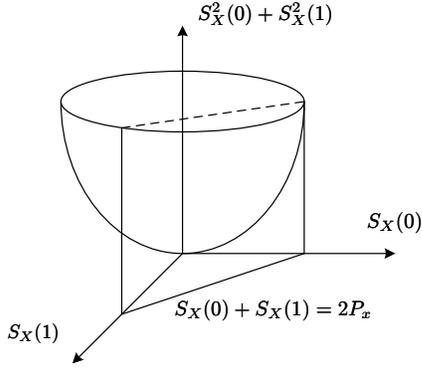,width=2.3in ,
clip=true}} \caption{\footnotesize{A geometric illustration of the
non-convex optimization problem formulated in
(\ref{eqn:H0-nonconvex}).}} \label{fig:nonconvex}
\end{figure}

Here we shall use the difference between $\mathbb{E}[T_{n,0}]$ and $\mathbb{E}[T_{n,1}]$ to determine the detection performance.
Suppose that we can control the spectral mask $\{S_X^{(n)}(k)\}$ of the
transmitted signal, we would like to maximize the difference
between $\mathbb{E}[T_{n,0}]$ and $\mathbb{E}[T_{n,1}]$, i.e.,
\begin{align}
&\mathrm{maximize}  \ \ \mathbb{E}[T_{n,1}]-\mathbb{E}[T_{n,0}]
\nonumber \\
&\  \ \ \ \ \mathrm{s.t.} \ \  \ \  \frac{1}{n}\sum_{k=0}^{n-1} S_X^{(n)}(k) = P_x \nonumber \\
&\ \ \ \ \ \ \  \ \ \ \  \ \  \ S_X^{(n)}(k) \geq 0, \ \ \  k=0, 1,
\ldots, n-1
\end{align}
with the optimization variables $\{S_X^{(n)}(k)\}_{k=0}^{n-1}$. For large $n$, this
problem is equivalent to
\begin{align}\label{eqn:H0-nonconvex}
&\mathrm{maximize}  \ \ \sum_{k=0}^{n-1} \left[S_X^{(n)}(k)\right]^2
\nonumber \\
&\ \ \ \ \ \mathrm{s.t.} \  \ \ \ \ \  \sum_{k=0}^{n-1} S_X^{(n)}(k) =n P_x \nonumber \\
&\ \ \ \ \  \ \ \ \ \ \ \ \ \ \ S_X^{(n)}(k) \geq 0, \ \ \  k=0, 1,
\ldots, n-1
\end{align}
which maximizes a convex function over a hyperplane. In the
sequel, we will show how to solve this nonconvex optimization
problem.

To solve (\ref{eqn:H0-nonconvex}), we first look at its
geometrical representation, as shown in Fig. \ref{fig:nonconvex}.
It is easy to see that optimal solutions fall into the
intersection of the convex surface of the objective function and
the hyperplane determined by the constraints. Thus, the optimal
solutions are given by
\begin{equation}\label{eqn:H0-nonconvex-optimal}
\left\{%
\begin{array}{ll}
     S_X^{(n)}(j) = n P_x, & j \in \{0, 1, \ldots, n-1\} \\
     S_X^{(n)}(k)=0, & 0 \leq k \leq n-1,\ \mathrm{and}\ k \neq j \\
\end{array}%
\right.
\end{equation}
for any arbitrary $j$, implying that all the transmit power is concentrated in a single
frequency bin. Accordingly, the optimal value of
(\ref{eqn:H0-nonconvex}) is
\begin{equation}
\mathbb{E}[T_{n,1}]-\mathbb{E}[T_{n,0}] = n^2 P_x^2.
\end{equation}

On the other hand, the worst case occurs when
\begin{equation}
S_X^{(n)}(k) = P_x, \ \ \ \ \ \ \ k=0, 1, \ldots, n-1
\end{equation}
which makes $\mathbb{E}[T_{n,1}]-\mathbb{E}[T_{n,0}] = nP_x^2$. In this extreme case, the
spectral mask function is flat across the spectrum. A
representative example of such a spectral feature is the white
noise-like signal. This result is consistent with our intuition
since it is generally difficult to distinguish a white Gaussian
signal from additive white Gaussian noise.  The optimal detector
for such a case is the energy detector (radiometer)
\cite{Poor1994}\cite{Kay98II} provided that the noise power is
perfectly known.

\section{Simulation Results}\label{sec:sim}

In this section, we numerically evaluate the proposed spectra
feature correlation-based spectrum sensing algorithm for both ATSC
and NTSC signals.

First, we obtain the ``clean'' baseband TV signals by capturing
the TV signals in the RF front-end and then transforming the
signals from the ultra-high frequency (UHF) bands to the baseband.
Through processing, the TV signals in the baseband are sampled at
a rate $6\times 10^6$ samples/sec, with $6$ MHz bandwidth. The TV
data are divided into a number of blocks, each of which is $6$
msecs. The spectral features are obtained by computing the (averaged)
periodograms of the clean TV signals.

We study the sensing performance for two channel models: AWGN and
multipath fading. For the multipath fading channel model, we apply
the ITU Pedestrian B model, whose power delay profile is given in
Table \ref{table:ex1}. The root mean square (RMS) delay spread of
the ITU Ped-B model is $633$ \emph{ns}. We pass the clean TV
signals through the channel models to simulate the real wireless
propagation environment.

\begin{table}
\renewcommand{\arraystretch}{1.3}
\caption{ITU Pedestrian-B Multipath Channel - Power Delay Profile}
\label{table:ex1} \centering
\begin{tabular}{|c||c|c|c|c|c|c|c}
\hline
\bfseries $\mathrm{Delays\ (ns)}$  & 0 &  200  &  800  &  1200 & 2300 & 3700    \\
\hline
\bfseries $\mathrm{Avg.\ Power Gain\ (dB)}$  & 0 &  -0.9 &  -4.9 &  -8.0 & -7.8 & -23.9      \\
\hline
\end{tabular}
\end{table}

\begin{figure}[t]
\centering{\epsfig{figure=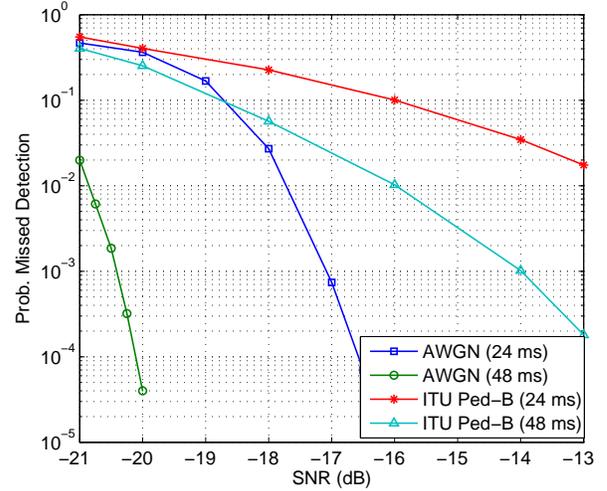,width=3.4in,
clip=true}}
\caption{\footnotesize{The missed detection rate of the proposed
spectrum sensing algorithm for ATSC signals, with a false alarm
rate less than 0.001. The detection interval is 24 \emph{ms}.}}
\label{fig:atsc-sim}
\end{figure}

\begin{figure}[t]
\centering{\epsfig{figure=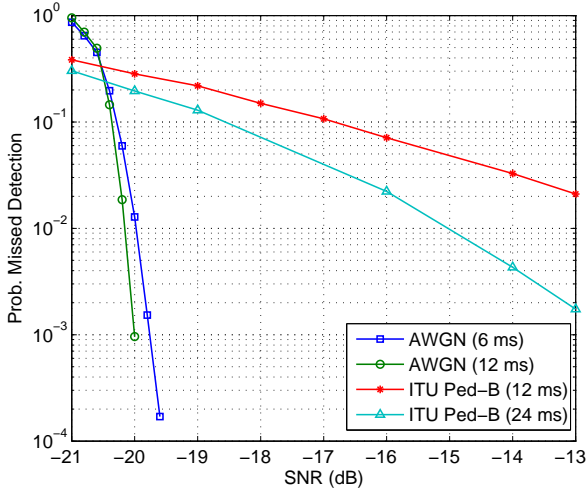,width=3.4in,
clip=true}}
\caption{\footnotesize{The missed detection rate of the proposed
spectrum sensing algorithm for NTSC signals, with a false alarm
rate less than 0.001. The results for AWGN channels are not shown
because they are lower than $10^{-4}$ in the above SNR regions.
The detection interval is 24 \emph{ms}.}} \label{fig:ntsc-sim}
\end{figure}

For both ATSC and NTSC signals, we choose the test thresholds such
that their false alarm rates are less than 0.001. We
use the white Gaussian noise to test the sensing algorithms to
make sure the false alarm rates are less than 0.001. Once we find
the test thresholds for ATSC and NTSC signals, we can simulate and calculate
the missed detection rates. For each SNR value, we simulate the
sensing algorithms for 25,000 realizations. The
simulation results for ATSC and NTSC are plotted in Figs.
\ref{fig:atsc-sim} and \ref{fig:ntsc-sim}.
%

For both ATSC and NTSC signals, the spectral feature detector can
reliably detect the signals at SNR$=-20$ dB with a missed
detection rate less than $0.01$ in the AWGN channels. It can be
observed that the NTSC signal is easier to detect than the ATSC
signal since the NTSC signal has three sharp spectral features
while the ATSC contains a large amount of flat spectrum and has only one feature corresponding to the pilot. This observation
is consistent with our optimization analysis in Section
\ref{sec:opt}.

\section{Conclusion}\label{sec:cls}

In this paper, we have proposed a spectral feature detector for
spectrum sensing in CR networks. The basic strategy is to use the
correlation between the periodogram of the received signal and the
\emph{a priori} spectral features. Using the asymptotic properties
of Toeplitz and circular matrices, we have shown that this
spectral feature detector is asymptotically optimal at very low
SNR and with a large block size. In addition, we have performed
optimization analysis on the effects of spectral features on the
sensing performance. The analytical results show that the signals
with sharp spectral features are easier to detect compared with
those with relatively flat spectra. The simulation results show
that the proposed spectral feature correlation detector can
reliably detect analog and digital TV signals at SNR as low as
$-20$ dB.


\useRomanappendicesfalse
\appendices

\section{Proof of Theorem \ref{thm1}}\label{appendix:a}

From the definition of the two test statistics, we have
\begin{equation}
 \lim_{n\rightarrow \infty} \left| T_{\mathrm{LRT}, n}-T_n
\right| = \lim_{n\rightarrow \infty}\frac{1}{n} \left|\mathbf{y}^*
\boldsymbol{\Sigma}_n \mathbf{y}- \mathbf{y}^* W_n^* \Lambda_n
 W_n \mathbf{y}
 \right|
\end{equation}
where
\begin{equation}
\Lambda_n= \left(%
\begin{array}{cccc}
   S_X^{(n)}(0) &  &  & 0 \\
   \vdots & \ddots &  & \vdots \\
    &  &  &  \\
   0 & \cdots& &  S_X^{(n)}(n-1) \\
\end{array}%
\right)
\end{equation}
is a diagonal matrix with the PSD of the incumbent signal in the
diagonal, and $W_n$ is the DFT matrix defined as
\begin{equation}
W_n = 
\left[%
\begin{array}{ccccc}
  1 & 1 & 1 & \cdots & 1 \\
  1 & w_n & w_n^2 & \cdots & w_n^{n-1} \\
  1 & w_n^2 & w_n^4 & \cdots & w_n^{2(n-1)} \\
  \vdots & \vdots & \vdots & & \vdots \\
  1 & w_n^{n-1} & w_n^{2(n-1)} & \cdots & w_n^{(n-1)(n-1)}\\
\end{array}%
\right]
\end{equation}
with $w_n=e^{-j2\pi/n}$ being a primitive $n$th root of unity.
Consequently,
\begin{align}\label{eqn:app-diff}
 \lim_{n\rightarrow \infty} \left| T_{\mathrm{LRT}, n}-T_n
\right| &= \lim_{n\rightarrow \infty}\frac{1}{n}
\left|\mathbf{y}^* \left(\boldsymbol{\Sigma}_n - W_n^* \Lambda_n
 W_n \right) \mathbf{y}
 \right|  \nonumber\\
 &= \lim_{n\rightarrow \infty}\frac{1}{n} \left|\mathbf{y}^*
\left(\boldsymbol{\Sigma}_n - C_n \right) \mathbf{y} \right|
\end{align}
where $C_n  \stackrel{\Delta} {=} W_n^* \Lambda_n
 W_n$ is a circular matrix. It has been shown in \cite{Gray2006} that
the Toepliz matrix $\boldsymbol{\Sigma}_n$ is asymptotically
equivalent to the circular matrix $C_n$ since the weak norm
(Hilbert-Schmidt norm) of $\boldsymbol{\Sigma}_n-C_n$ goes to zero
\cite{Gray2006}, i.e.,
\begin{equation}
 \lim_{n\rightarrow \infty} \|\boldsymbol{\Sigma}_n -C_n  \| = 0.
\end{equation}
Thus, we can establish (\ref{eqn:theorem1}) from (\ref{eqn:app-diff}).

%

\bibliographystyle{IEEEtran}
\bibliography{IEEEabrv,ref}

\end{document}